# Atomic mechanism of phase transition between metallic and semiconducting MoS$_2$ single-layers

Yung-Chang Lin,[1] Dumitru O. Dumcenco,[2] Ying-Sheng Huang,[2] Kazu Suenaga[1*]

[1]National Institute of Advanced Industrial Science and Technology (AIST), Tsukuba 305-8565, Japan

[2]Department of Electronic Engineering, National Taiwan University of Science and Technology, Taipei 10607, Taiwan

Structural transformation between metallic (1T) and semiconducting (2H) phases of single-layered MoS$_2$ was systematically investigated by an *in situ* STEM with atomic precision. The 1T/2H phase transition is comprised of S and/or Mo atomic-plane glides, and requires an intermediate phase (α-phase) as an indispensable precursor. Migration of two kinds of boundaries (β and γ-boundaries) is also found to be responsible for the growth of the second phase. The 1T phase can be intentionally introduced in the 2H matrix by using a high dose of incident electron beam during heating the MoS$_2$ single-layers up to 400~700°C in high vacuum and indeed controllable in size. This work may lead to the possible fabrication of composite nano-devices made of local domains with distinct electronic properties.

KEYWORDS: MoS$_2$, phase transition, intra-layer lubricate, in situ observations, low-dimensional materials

Phase transition in a thermodynamic system is one of the most fundamental phenomena and of great technological importance in material science because the properties can be altered without any extra additional atom. Phase transition in a solid undergoes with collective atomic displacement, and such atomic process has been so far investigated by macroscopic viewpoints. In order to deeply understand how the new phase nucleates and then grows from the initial phase, *in situ* observation during the phase transition is definitively required with atomic resolution. Especially identifying the atomic structures of the nucleate point and the boundary of the transition frontier is quite prerequisite for energetic considerations of the phase transitions.

$MoS_2$ is widely used as a practical solid lubricant since its discovery in 1960s[1,2]. The crystal is built up of atomic layers stacking by van der Waals force, and each layer is constructed of strong in-plane bonding of S-Mo-S' triple atomic planes. Recently, single-layered molybdenum disulfide (SL-$MoS_2$), a direct bandgap quasi-2D semiconductor, has shown its great potential for applications in electrical and optoelectronic devices[3-5]. Interestingly, one of the unique features of $MoS_2$ is the polymorphism with distinct electronic characteristics. Depending on the arrangement of S atoms, SL-$MoS_2$ appears in two distinct symmetry: 2H (trigonal prismatic $D_{3h}$) and 1T (octahedral $O_h$) phases (Fig. 1a and 1b). The two phases are supposed to exhibit completely different electronic structures where 2H phase is semiconducting while 1T is metallic[6-8]. Both phases can easily convert to each other through an intra-layer atomic plane glide, which involves a transversal displacement of one of the S-planes. The 1T phase was first reported to transform from 2H-$MoS_2$ by Li and K intercalation[6,9], while it is also known to be stabilized by substitutional doping of Re, Tc, and Mn atoms which serve as electron donor[10]. The phase transitions between 2H and 1T or 2H' phases due to the atomic plane gliding are presented in Fig. 1c and 1d.

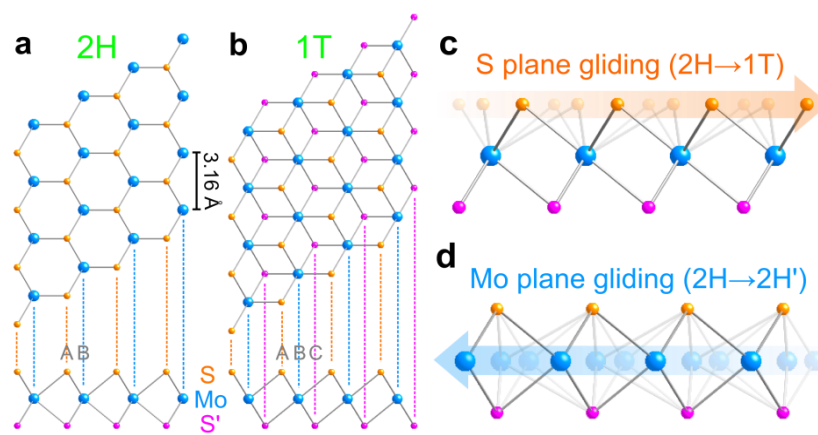

**Figure 1. Polymorphs of SL-MoS$_2$.** The schematic models of SL-MoS$_2$ with 2H (**a**) and 1T (**b**) phases in basal plane and cross section view. Atom color: Mo, blue; top S, orange; bottom S', purple. The incident electron beam (e-beam) transmits from top to bottom. The 2H phase shows hexagon lattice with threefold symmetry and the atomic stacking sequence (S-Mo-S') of ABA. The 1T phase shows atomic stacking sequence (S-Mo-S') of ABC where the bottom S' plane occupy the hollow center (HC) of 2H hexagonal lattice. **c,** S plane glides in a distance equivalent to $a/\sqrt{3}$ ($a = 3.16$ Å) and occupies the HC site of 2H hexagon, which results in 2H→1T phase transition. **d,** Mo plane gliding results in 2H→2H' transition. The three planes (Mo, S, and S') in single layer MoS$_2$ can glide individually to perform different transitions.

Although the coexistence of metallic and semiconducting phases was indeed reported in chemically exfoliated MoS$_2$ by Eda *et al*[11], the actual dynamical process of the transformation between 2H and 1T phases involving the intra-layer atomic plane gliding has never been experimentally proved nor the atomic process of phase transition investigated *in situ*. When one consider the possibilities to intentionally introduce the phase transition in single-layered materials with a controllable manner, the atomic process of phase transition and its boundary structures must be corroborated in order to design future low-dimensional devices.

We demonstrate here *in-situ* observations of transformation process between 2H and 1T phases in SL-MoS$_2$ at high temperatures. In order to monitor the phase transition *in situ*, we operated an aberration-corrected STEM at 60 kV to visualize the dynamic process of the atomic motions in SL-MoS$_2$. This technique has been already used and verified for studying another ideal two-dimensional (2D) material, graphene, with dislocations[12,13], grain boundaries[14,15], and the dynamics of the defect movements[16,17]. In case of MoS$_2$, the number of studies is still limited except ones that studied defects and native grain boundary between two MoS$_2$ domains[18,19].

The MoS$_2$ specimen doped with 0.6 at% Re was exfoliated and transferred to a micro-grid[20,21]. In order to promote the phase transition, the specimen was heated up to 400~700 degree in a microscope, which is close to the thermodynamically equilibrium temperature. An example of the phase transition seen in sequential annular dark field (ADF) images is shown in Figs. 2a-2d in which a step-by-step progress of MoS$_2$ phase transformation at T=600°C is represented (Movie M1). Figs. 2e-2h show the schematics correlating with the ADF images from Figs. 2a-2h to illustrate the structure changes in the MoS$_2$ lattice. The Re dopants (indicated by arrows in Fig. 2a) tend to substitute at the Mo sites and display brighter contrast[21]. The initial MoS$_2$ lattice (Fig. 2a) exhibits the 2H phase that the honeycomb structure consist of 3 Mo atoms and 3 overlapped S pairs in a hexagon. At t = 100s, two identical band-like structures (labeled as "α") in Fig. 2b gradually form along two zigzag directions. This α-phase is a precursor which basically consists of 3~4 constricted MoS$_2$ zigzag chains with ~33% higher proportion of Mo and S atoms.

When two non-parallel α-phase are being contacted, the number of atoms at the local acute angle corner becomes very dense that triggers the atoms gliding towards the less atomic concentration area to release the stress. A triangular shape of 1T phase forms in Fig. 2c. The 1T phase showing a different S contrast can be unambiguously discriminated from the 2H in ADF image (Fig. S1). After continuous e-beam scanning, the size of the 1T phase can be gradually enlarged to ~8.47 nm$^2$ (Fig. 2d). Two new phase boundaries (β and γ in Fig. 2d) are found at the edges of 1T phase. The structures and dynamic behaviors of these boundaries will be discussed in detail below. The phase transformation happens only at the e-beam scanning area.

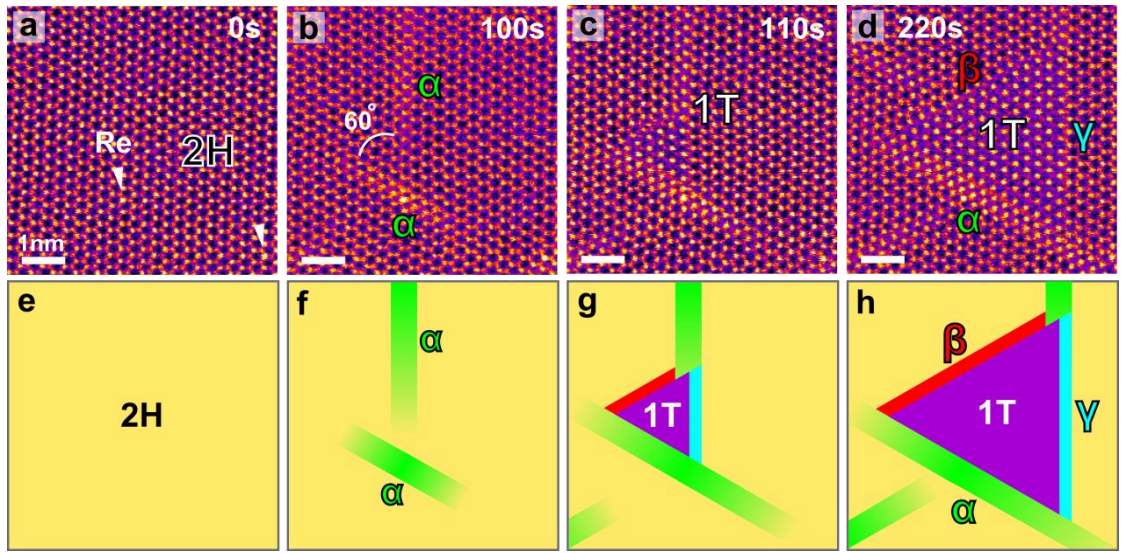

**Figure 2. Evolution of 2H → 1T phase transformation in SL-MoS$_2$ at T=600°C.**

**a**, SL-MoS$_2$ doped with Re substitution dopants (pointed by arrows) shows the initial 2H phase of a hexagonal lattice structure with a clear HC. **b**, At t = 100s, two identical intermediate (precursor) phases denoted as "α" form with an angle of 60°, which consists of three constricted Mo zigzag chains. **c**, A triangle shape of 1T phase (~1.08 nm$^2$) appears at the acute corner between two α-phases at t = 110s. The 1T phase shows noticeable contrast due to the S atoms at HC sites (Fig. S1). **d**, At t = 220s, the size of transformed 1T phase becomes enlarged to ~8.47 nm$^2$. Three different boundaries (α, β, and γ) are found at three edges between the 1T and 2H phases. **e-h**, Simple schematics illustration of 2H→1T phase transition following the ADF images in **a-d**, respectively.

The phase transformation in SL-MoS$_2$ involves numerous atomic displacements besides the simple atomic plane gliding. We have investigated the atomic process of the phase transitions on more than 100 cases by *in situ* STEM. The cases involve the phase transitions of 2H to 1T, 1T to 2H, 2H to 2H' and 1T to 1T' (Table 1). Note that 2H' is a 60 degree (or 180 degree) rotational phase of 2H. We have then tried to categorize these transitions into the three important elemental steps namely the nucleation (formation) of the α-phase as a precursor or an intermediate state as well as the migration of β and γ boundaries. Figure 3 shows these three scenarios with independent sequential ADF images.

**Scenario I: α-phase formation**

The α-phase is a precursor structure which is indispensable in prior to the phase transition. This is an intermediate state but is thermodynamically stable. In the α-phase, the Mo atoms do not show the trigonal arrangements but align as the zigzag chains. The Mo-Mo distance is substantially shrunk by ~15% as the result of in-plane constriction (Fig. 3a, Movie M2). The Mo-Mo in-plane constriction is probably connected to the S out-of-plane displacement induced by e-beams. The local strains must be released by changing the Mo-S bond angles although our STEM observation is not capable to prove the exact strained structure of this α-phase. Fig. 3d shows a schematic of in-plane constriction (green arrows in left) and structure model of the α-phase (right). The S out-of-plane displacement can propagate in the zigzag direction and so elongates the α-phase (Fig. 3b). Interestingly, the α-phase has a strong tendency to nucleate at the vicinity of Re substitution dopants. Assumable the initial out-of-plane protuberance of Re-S bond could help the S out-of-plane displacement and thus the formation of α-phase[21]. Note here that, the α-phase always consists of three or four MoS$_2$ zigzag chains and not expand in width (Fig. S2, Movie M3). There is no atom loss during the formation and elongation (directional growth) of the α-phase. In Fig. 3c, during the migration of the central α-phase, the left part of the bottom α-phase indeed disappears and recovers back to the original 2H structure (pointed by arrow). Such a reversible phase transformation between 2H and α-phase does prove that there is no massive atom loss during the α-phase formation. Even though the e-beam is definitely required to displace the S atoms out-of-plane, no atom is kicked out by the knock-on effect. In Supplementary information, we show another scenario in which the prolonged e-beam irradiation occasionally leads to the loss of a MoS$_2$ zigzag chain and form an agglomerated structure on the MoS$_2$ surface (Fig. S3, Movie M4). The α-phase hardly forms at room temperatures.

**Scenario II: β boundary migration**

The β-boundary is a twin boundary containing the Mo-S four-membered rings which was in principle the same as the one recently observed at the boundary of two 60 degree rotated MoS$_2$ domains (i.e., 2H and 2H') synthesized by chemical vapor deposition[19]. The S atoms in the β-boundary are four-coordinated despite all the other phases have the three-coordinated S atoms. The 2H to 2H' phase transition requires Mo plane gliding and generates β-boundary (Fig. S4, Movie M5). The β-boundary can also be found between 2H and 1T phases (Fig. 1d), in this case, it becomes no longer a twin boundary. Here the latter case is presented as an example. Fig. 3e shows a typical ADF image of MoS$_2$ in 2H phase with a threefold symmetry and the orientation is described by a blue triangle. At t = 60s, β-boundary (highlighted by yellow ribbon) appears in middle of 2H-MoS$_2$ (Fig. 3f). The left-hand side of the β-boundary becomes the 1T phase. Note that the β-boundary shows up when a Mo-plane and an S-plane are both glided during the 2H to 1T phase transition. Another simpler transition from 2H to 1T with only one S-plane glide results in the formation of γ-boundary between two phases (Table 1 and Movie M8). Then, the Mo + S(or S') atoms gliding across the β-boundary (Fig. 3h, left) drives the β-boundary to migrate (Fig. 3h, right). Fig. 3g shows the β-boundary migrates rightward at t = 110s. Detail structure and gliding model are also described in Fig. S6.

**Scenario III: γ boundary migration**

The γ-boundary consists of two constricted $MoS_2$ zigzag chains. While the α-phase is made of exclusively three or four $MoS_2$ zigzag chains, the γ-boundary has always two chains. The S atoms at γ-boundary remain three-coordinated to the Mo atoms. Fig. 3i shows an ADF image of $MoS_2$ in the initial 1T phase. The two S planes are vertically misaligned in 1T phase. At t = 20s, γ-boundary (highlighted by purple ribbon) appears in between of initial 1T phase and nucleated 2H phase (Fig. 3j). The left-hand side of the γ-boundary transforms from 1T to 2H phase by an S plane gliding. The schematic model shown in Fig. 3l (left) illustrates the S atoms sequentially glide towards γ-boundary, and then the 2H phase increases with the γ-boundary migration (Fig. 3l, right). The γ-boundary migrates rightward step-by-step consecutively and also a non-straight boundary showing the atomic step is pointed by arrow (Fig. 3k, Movie M6). Detail structure and gliding model are described in Fig. S6.

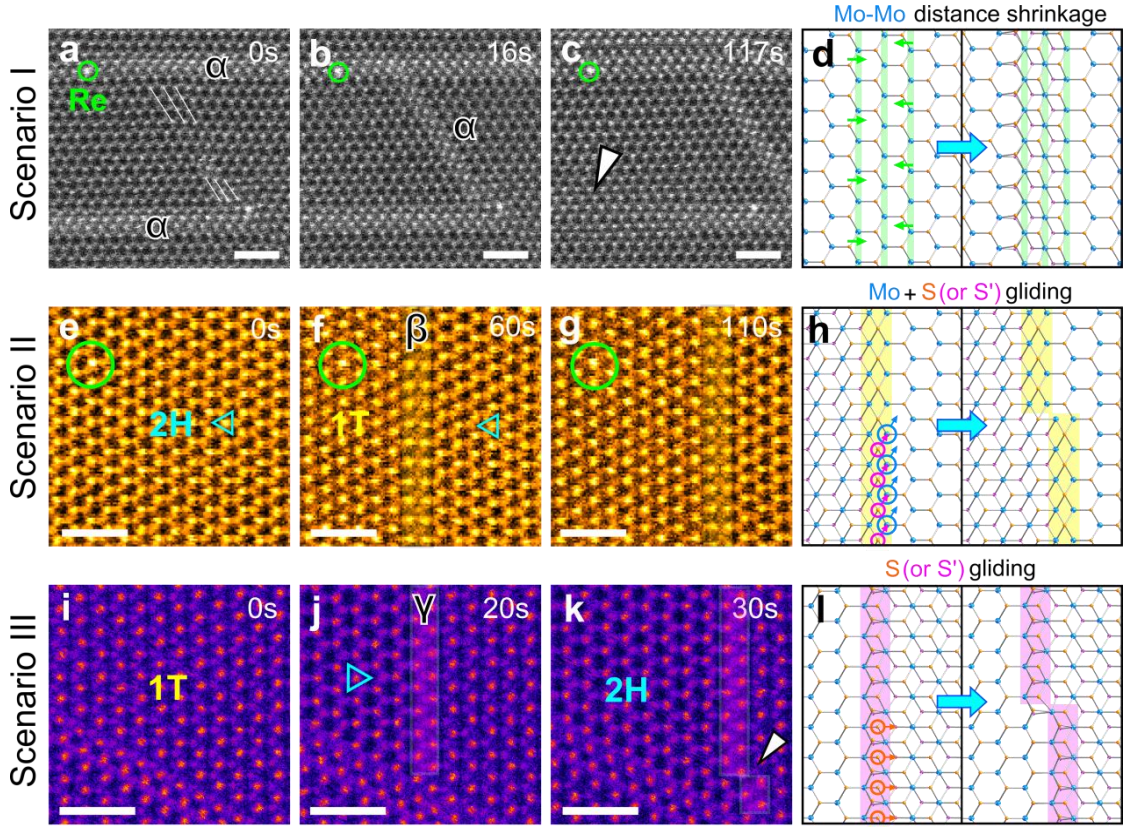

**Figure 3. Three elemental steps responsible for phase transitions in SL-MoS$_2$ (T = 600°C).**

**Scenario I: α-phase (three or four zigzag chains) formation. a**, Nucleation of α-phase in the angle of 60 degree with the other α-phases. The α-phase shows three or four constricted zigzag MoS$_2$ chains. Three white lines guide the distance between zigzag chains where the in-plane constriction in α-phase is ~15% of the original MoS$_2$. **b**, Growth of α-phase. **c**, At t = 117s, the α-phase starts to migrate rightward. The left side of the bottom α-phase (pointed by arrow) disappears and recovers to initial MoS$_2$ lattice. Re dopants are marked by green circles as markers. **d**, Constrain (green arrows) induces the strain in-plane (left) and the model α-phase forms with shrunk Mo-Mo distance forms (right). The S atoms in the α-phase are also vertically misaligned. **Scenario II: β-boundary (4-6-6-4) migration. e**, SL-MoS$_2$ with 2H phase. The orientation of initial 2H phase is determined by the blue triangle. **f**, At t = 60s, β-boundary (highlighted by yellow ribbon) appears in middle of 2H-MoS$_2$. The left-hand side of β-boundary shows 1T phase. **g**, The β-boundary migrates rightward and 1T phase is enlarged. **h**, The schematic model of before (top) and after (bottom) the Mo + S(or S') atoms gliding which causes the β-boundary migration. **Scenario III: γ-boundary (two zigzag chains) migration. i**, SL-MoS2 with 1T phase (α-phase is also visible). **j**, At t = 20s, γ-boundary (highlighted by purple ribbon) appears in the middle. The left-hand side of γ-boundary shows the nucleated 2H phase. **k**, The γ-boundary migrates rightward and shows a non-straight structure. **l**, The schematic model of before (top) and after (bottom) top S(or S') atoms gliding which drives the γ-boundary migration. Scale bar: 1 nm.

The distinct features of boundary structure are also corroborated by electron energy loss spectroscopy (EELS). Recent literatures reported single impurity atoms of Si in graphene lattice discriminated in the three and four coordinated configurations[22,23]. Therefore the bonding states of S atoms in the newly discovered boundaries are intriguing and would be most prominent in the S $L$-edge. Here the electron energy-loss near-edge structure (ENLES) for S $L_{23}$-edge and Mo $M_{45}$-edge was recorded on the β and γ boundary regions. Each image-spectrum consists of 12×12 pixels by using a 0.1 nm probe with 0.05 nm increments for each step. Each spectrum was acquired in 0.5 sec and is summed up in vertical direction to increase the SN ratio. Energy dispersion for the recorded spectra was 0.25 eV and the zero loss width of incident e-beam was about 0.35 eV.

Figs. 4a and 4c show the ADF images of β and γ boundaries overlapped with schematics of structure models, respectively. The spectrum in blue (Fig. 4b top) was taken from the normal threefold S atoms in the MoS$_2$ 2H lattice (Fig. 4a between two blue ribbons), in which the S bonded with three Mo atoms, denoted as S-Mo$_3$. The near-edge fine structure of S-Mo$_3$ presents sharp S $L_{2,3}$-edge at 165 and 178 eV as well as a broader peak at 193 eV (pointed by arrows). The red spectrum (Fig. 4b lower) was taken from the β-boundary in which the S atoms are supposedly fourfold (S-Mo$_4$) (Fig. 4a red ribbons). The second and third *L*-edge peak show significant decreases in the red spectrum. Especially the flattening of the second peak at 178 eV is quite prominent. The *L*-edge reflects the excitation of 2p electrons to unoccupied 3d states, therefore, the S-Mo$_4$ has distinct hybridization states rather than S-Mo$_3$. On the other hand, the green spectrum (Fig. 4d lower) taken from the γ-boundary with the misaligned threefold S atoms (green ribbon area in Fig. 4c) presents no significant difference in the *L*-edge spectrum. Similarly we do not find any considerable changes in S *L*-edge fine structures for α-phase, 2H phase, or 1T phase, which is suggestive that all the threefold S atoms shows similar *L*-edge even though the Mo-S-Mo bonds are slightly distorted.

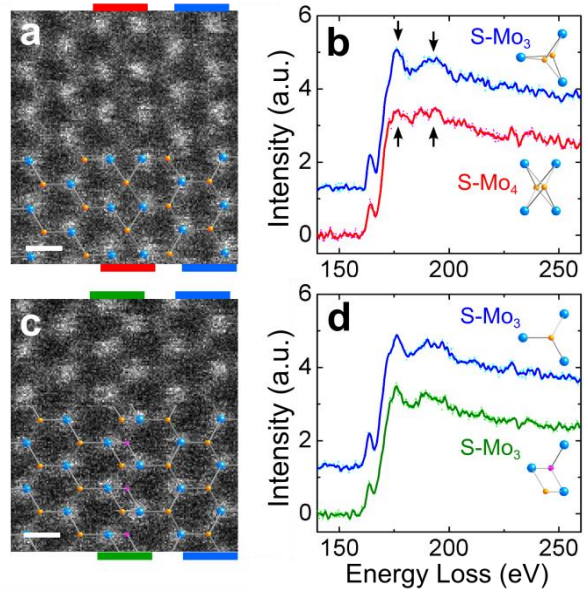

**Figure 4. EELS for fourfold and threefold coordinated S in β and γ-boundaries.**

**a**, ADF image and overlaid schematic model of β-boundary. The S atoms are threefold coordinated in MoS$_2$ 2H lattice (blue ribbon) but fourfold in β-boundary (red ribbon), where the EEL spectra were acquired. **b**, EEL spectra for *L*-edge of threefold S in MoS$_2$ (blue, top) and fourfold S in β-boundary (red, bottom). **c**, The ADF image and overlaid schematic model of γ-boundary. The S atoms are threefold coordinated in the γ-boundary (green ribbon) but misaligned in the c-axis. **d**, EEL spectra for *L*-edge of threefold S in MoS$_2$ (blue, top) and fourfold S in γ-boundary (green, bottom). The EEL spectra in **b** and **d** are smoothed based on the dotted data points. Scale bars: 0.25 nm.

The phase transformation presented here involves complicated dynamic processes. The gliding planes decide the relationship with the initial and final phases, as well as the correlated phase boundaries. Table 1 catalogs the result of systematic investigation of the discovered phase transformations in SL-MoS$_2$ and all the corresponding schematic model and detail discussion are presented in the supplementary information. In an attempt to control the phase transformation by e-beam, we have continuously recorded the size of transformed area as a function of time. The data points plot in Fig. 5a shows the relation between the electron dose and the area of the transformed phase at different thermal environment (400°C<T<700°C). Here we use the dose instead of time because the data were normalized by the dose rate for the unit area. The phase transformation area (*A*) increases as $A \propto e^{\sigma(D-D_0)}$, where $D$ is the electron dose, $D_0 \approx 40$ MeV/nm$^2$ is the threshold before triggering the phase transformation, $\sigma \approx 0.028$~$0.061$. The relation between the transformed area and the electron dose can be divided into two regions by $D_0$. In the region (I), $D < D_0$, the phase transformation does not start until the electron dose create the intermediate state structures, the α-phases. In the region (II), $D > D_0$, the phase transformation starts to increase the size as the time being with the dose increase.

As the e-beam scanning area and the irradiation time can be easily controlled in an STEM, we can intentionally introduce the phase transition at a chosen area with a preferable size. Because the 1T and 2H phases have distinct electronic properties, the controllable local phase transition may lead to the bottom up processes to fabricate nano-electronics. To explore these possibilities, we demonstrate in Fig. 5 several attempts to produce prototypes of nano-devices. Fig. 5b shows a serial junction of semiconductor and metallic phase which can be regarded as a Schottky diode. A local semiconductor region sandwiched two metallic electrodes is as a nano-scale transistor (Fig. 5c). A metallic wire can be embedded in the semiconducting matrix as a quantum lead (Fig. 5d). Finally, quantum dots (but in a triangular shape) can be stably produced in the initial phase; a metallic quantum dot embedded in semiconducting (Fig. 5e), or vice versa (Fig. 5f).

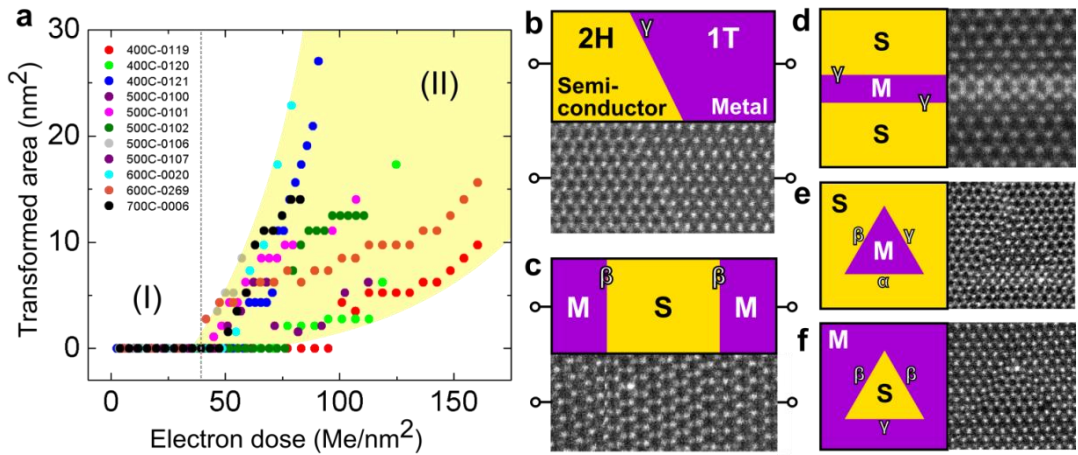

**Figure 5. Time-dependence of phase transformation process and fabrication of nano-devices in SL-MoS$_2$.**

**a**, The area of transformed phase as a function of total electron dose (instead of time). The dispersion is divided into two regions by the threshold electron dose (grey dotted line). The region (I) involves the initial step to create the intermediate structures, α-phase. In region (II), the phase transition initiates and the transformed area increases with the increasing electron doses as the time being. The experiment data sets were recorded at various temperatures ranging from 400℃ to 700℃. **b-f**, Attempts to create the prototypes of nano-devices. **b**, 2H and 1T heterostructure with a γ-boundary as a Schottky diode. **c**, 2H sandwiched in two 1T phases with two β-boundaries as a Schottky barrier nano-transistor. **d**, Single Mo hexagon chain formed on top of the 2H-$MoS_2$ as a Metallic quantum wire. **e**, 1T phase embedded in 2H phase as an embedded metallic quantum dot. **f**, 2H phase embedded in 1T phase an embedded semiconducting quantum dot.

| Table 1 \| Summary of phase transformation in SL-$MoS_2$ | | |
|---|---|---|
| **Phase transition** | **Gliding plane** | **Boundary structures** |
| 2H → 1T | Mo + S(or S') | α + β + γ |
|  | S(or S') | α + γ   or   *γ |
| 2H → 2H' | Mo | α + β + γ |
|  | S + S' | *γ |
| 1T → 2H | Mo + S(or S') | α + β + γ |
|  | S(or S') | α + γ |
| 1T → 1T' | † Mo + S(or S') | † α + β + γ |

* The special case that no α phase formed before phase transition (Fig. S5 and Movie M7, M8).
† The structure that we have not found yet.

These structures were so far fabricated only in an electron microscope and their functions have not yet been experimentally confirmed. The transfer process and the preventing surface contaminations are still obstacles to go beyond. The relatively stable single-layered structures of this system are however very promising to get the first electronic device which is made of single-layered only.

Though the low-dimensional nano-device was first proposed using nanotubes of metallic and semiconducting components, it has been however turned out to difficult realize the nanotube composites with controlled chiralities. Patterning single-layer is definitely a more promising approach toward the realization of nano-devices. Similar phenomena for the MoWS$_2$ alloys with the tunable bandgaps[24] and for the n and p type doped dichalcogenides[21] would be extremely intriguing.

## Acknowledgements

The authors from AIST acknowledge the support from JST Research Acceleration Programme. D.O.D. and Y.S.H. acknowledge the support of the National Science Council of Taiwan under Projects NSC 100-2112-M-011-001-MY3 and NSC 101-2811-M-011-002.

## Author contributions

YCL performed experiments and analyzed data. DOD and YSH grew materials. KS and YCL designed experiments. YCL and KS co-wrote paper.

## Additional information

The authors declare no competing financial interests. Supplementary information accompanies this paper at www.nature.com/naturenanotechnology. Reprints and permission information is available online at http://www.nature.com/reprints. Correspondence and requests for materials should be addressed to K.S.

# References

1. Winer, W. O. Molybdenum disulphide as a lubricant: A review of the fundamental knowledge. *Wear* **10**, 422-452 (1967).

2. Holinski, R. & Gansheimer, J. A study of the lubricating mechanism of molybdenum disulphide. *Wear* **19**, 329-342 (1972).

3. Radisavljevic, B., Radenovic, A., Brivio, J., Giacometti, V. & Kis, A. Single-layer $MoS_2$ transistors. *Nat. Nanotech.* **6**, 147-150 (2011).

4. Wang, Q. H., Kalantar-Zadeh, K., Kis, A., Coleman, J. N. & Strano M. S. Electronics and optoelectronics of two-dimensional transition metal dichalcogenides. *Nat. Nanotech.* **7**, 699-712 (2012).

5. Chhowalla, M. *et al.* The chemistry of two-dimensional layered transition metal dichalcogenidenanosheets. Nat. Chem. **5**, 263-275 (2013).

6. Matthesis, L. F. Band structure of transition-metal-dichalcogenide layer compounds. *Phys. Rev. B* **8**, 3719-3740 (1973)

7. Wypych, F. & Schöllhorn, R. 1T-$MoS_2$, a new metallic modification of molybdenum disulfide. *J. Chem. Soc., Chem. Commun.* 1386-1388 (1992).

8. Bissessur, R., Kanatzidis, M. G., Schindler, J. L. & Kannewurf, C. R. Encapsulation of polymers into $MoS_2$ and metal to insulator transition in metastable $MoS_2$. *J. Chem. Soc., Chem. Commun.* 1582-1585 (1993).

9. Py, M. A. & Haering, R. R. Structural destabilization induced by Lithium intercalation in $MoS_2$ and related-compounds. *Can. J. Phys.* **61**, 76-84 (1983).

10. Enyashin, A. N. *et al.* New route for stabilization of 1T-$WS_2$ and $MoS_2$ phases. *J. Phys. Chem. C.* **115**, 24586-24591 (2011).

11. Eda, G. *et al.* Coherent atomic and electronic heterostructures of single-layer $MoS_2$. *ACS Nano* **6**, 7311-7317 (2012).

12. Hashimoto, A., Suenaga, K., Gloter, A., Urtia, K. & Iijima. S. Direct evidence for

atomic defects in graphene layers. *Nature* **430**, 870-873 (2004).

13. Kotakoski, J., Krasheninnikov, A. V., Kaiser, U. & Meyer, J. C. From point defects in graphene to two-dimensional amorphous carbon. *Phys. Rev. Lett.* **106**, 105505 (2011).

14. Huang, P. Y. *et al.* Grains and grain boundaries in single-layer graphene atomic patchwork quilts. *Nature* **469**, 389-392 (2011).

15. Kurasch, S. *et al.* Atom-by-atom observation of grain boundary migration in graphene. Nano Lett. **12**, 3168-3173 (2012).

16. Warner, J. H. *et al.* Dislocation-driven deformations in Graphene. *Science* **337**, 209-212 (2012).

17. Lehitnen, O., Kurasch, S., Krasheninnikov, A. V. & Kaiser. U. Atomic scale study of the life cycle of a dislocation in graphene from birth to annihilation. *Nat. Comm.* **4**, 2098 (2013).

18. van der Zande, A. M. *et al.* Grains and grain boundaries in highly crystalline monolayer molybdenum disulfide. *Nat. Mater.* **12**, 554-561 (2013).

19. Zhou, W. *et al.* Intrinsic structural defects in monolayer molybdenum disulfide. *Nano Lett.* **13**, 2615-2622 (2013).

20. Tiong, K. K., Huang, Y. S. & Ho, C. H. Electrical and optical anisotropic properties of rhenium-doped molybdenum disulphide. *J. Alloys Compd.* **317-318**, 208-212 (2001).

21. Lin, Y. C. *et al.* Properties of individual dopant atoms in single-layer $MoS_2$: Atomic structure, migration, and enhanced reactivity. *Submitted*, (2013).

22. Zhou, W. *et al.* Direct determination of the chemical bonding of individual impurities in graphene. *Phys. Rev. Lett*. **109**, 206803 (2012).

23. Ramasse, Q. M. *et al.* Probing the bonding and electronic structure of single atom dopants in graphene with electron energy loss spectroscopy. *Nano Lett.* **DOI**:10.1021/nl304187e (2012).

24. Chen, Y. *et al.* Tunable band gap photoluminescence from atomically thin transition-metal dichalcogenide alloys. *ACS Nano* **5**, 4610-4616 (2013).

# Supplementary Information

# Atomic mechanism of phase transition between metallic and semiconducting MoS$_2$ single-layers

Yung-Chang Lin,[1] Dumitru O. Dumcenco,[2] Ying-Sheng Huang,[2] Kazu Suenaga[1*]

[1]National Institute of Advanced Industrial Science and Technology (AIST), Tsukuba 305-8565, Japan

[2]Department of Electronic Engineering, National Taiwan University of Science and Technology, Taipei 10607, Taiwan

**Content**

1. Identification of 2H an 1T phases by STEM-ADF image

2. Additional information of α-phase

    2.1. Expansion of α-phase area.

    2.2. Loss of a MoS$_2$ zigzag chain in α-phase.

3. Additional information of β and γ-boundaries

    3.1. β and γ-boundaries formed in 2H→2H' phase transition.

    3.2. 2H→2H' and 2H→1T' phase transitions without intermediate state (α-phase)

    3.3. Atomic gliding models for 2H↔2H' and 2H↔1T phase transitions.

1. **Identification of 2H and 1T phases by STEM-ADF image**

The ADF images of 2H-MoS$_2$ is quite straight forward to be interpreted because the two S planes are vertically overlapped (Fig. S1a) and give darker contrast in the HC. In the case of 1T-MoS$_2$, the Mo lattice remains the same as the 2H, while the two S planes are vertically misaligned (Fig. S1b). Ideally the two S atoms in different planes (S and S', orange and purple) give the same contrast in ADF image. Surprisingly, we found unambiguous contrast difference between the two S atoms in 1T phase. In Fig. S1c, the ADF image shows both 2H and 1T phases. Fig. S1d shows the ADF profile taken from the atoms along the armchair direction (indicated by rectangular in Fig. S1c). The contrasts due to S atoms are shown in the right-side of the ADF profile diagram in Fig. S1d. One S plane (S) displays a clear atomic profile, while the other (S') contributes relatively weaker contrast. The unexpected difference in the ADF contrast between two S planes (S and S') in 1T-MoS$_2$ can be originated from different reasons. One of the possibilities would be instability of the S' planes. Note that the bottom S plane (S') is supposed to have lower displacement threshold energy [1]. The other is an electron channeling effect. It would be interesting to further investigate the channeling effect in the materials among three atomic planes only. This asymmetric contrast in 1T phase is not coming from an inclined orientation of the specimen. As seen in Fig. S1c, the incident beam is completely normal in the 2H phase and so it is into the 1T phase nearby.

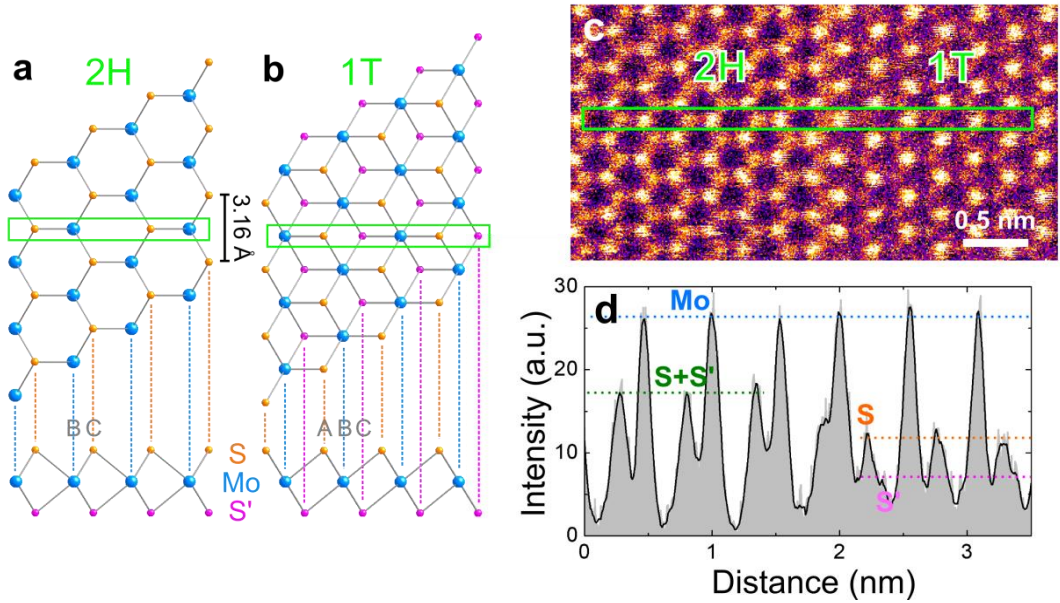

**Fig. S1.** The schematic models of SL-MoS2 with **(a)** 2H and **(b)** 1T phases. **(c)** The ADF image of SL-MoS$_2$ shows 2H phase (left) and 1T phase (right) with a γ-boundary in between. **(d)** The ADF profile taken from the rectangular area in **(c)**. Note that, the two S planes unexpectedly show up in distinct ADF contrast in 1T phase.

2. **Additional information of α-phase**

2.1. **Expansion of α-phase area.**

The α-phase consists of 3~4 zigzag chains (yellow lines in Fig. S2a) with a shrinkage Mo-Mo distance which displays like bright ribbons in the MoS$_2$ lattice. The α-phase is a stable structure which does not expand in width to the one containing more than 4 zigzag chains. Fig. S2 shows the sequential ADF images of the growth of α-phase and expand the coverage area by lining up in parallel. Fig. S2a shows a 2H-MoS$_2$ ribbon which consists of 4 hexagon chains (blue lines) is sandwiched by two α-phases. In Figs. S2b-S2d, another α-phase is growing in the 2H-MoS$_2$ ribbon from the bottom toward the top. Fig. S2e shows the final structure that three α-phases aligned side-by-side.

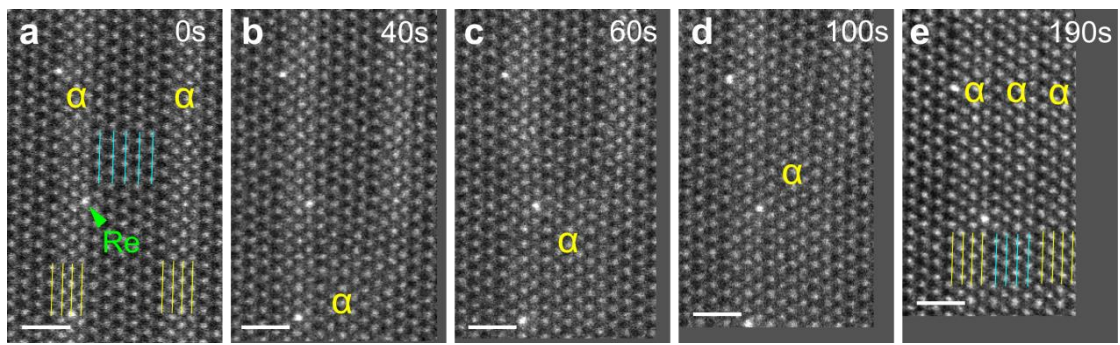

**Fig. S2.** Sequential ADF images of the α-phases growth side-by-side (Movie M3). Scale bar: 1nm.

### 2.2. Loss of a MoS$_2$ zigzag chain in α-phase.

Formation of α-phase is a reversible process and does not require the loss of atoms. However, the prolonged e-beam irradiation sometimes damage the α-phase. If the atoms in the α-phase are sputtered by e-beam, the α-phase will lose a chain of zigzag MoS$_2$ and recover to the original MoS$_2$ lattice phase. Figure S3 shows the sequential images illustrating the transition from α-phase to original 2H phase by losing atoms and form an agglomerated structure on the MoS$_2$ surface. See also for the Movie M4.

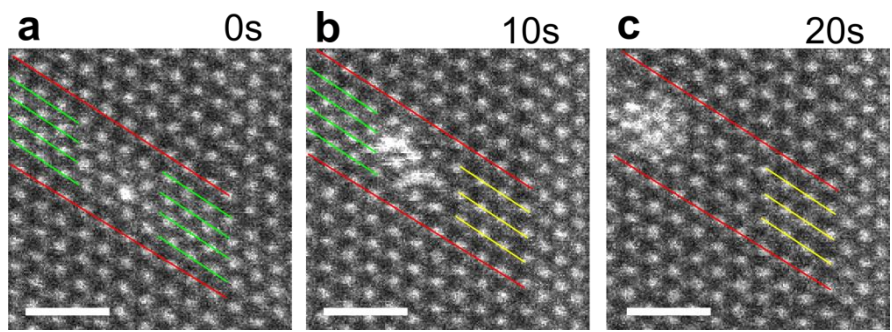

**Fig. S3.** (a) ADF image of a α-phase in 2H-MoS$_2$. The α-phase consists of four constricted MoS$_2$ zigzag chains indicated by green guide lines. The red line guides the eyes for zigzag edge of hexagons in 2H-MoS$_2$. **(b)** At t = 10s, the α-phase breaks by losing one zigzag chain and the structure returned to 2H phase (three yellow lines). (c) The lost zigzag MoS$_2$ atoms in α-phase agglomerate on the MoS$_2$ surface. Scale bar: 1 nm.

## 3. Additional information of β and γ-boundary

### 3.1. β and γ-boundary formed in 2H→2H' phase transition.

The phase transition between 2H and 2H' requires Mo plane gliding. Figs. S4a-4d show the sequential ADF images illustrating the evolution of 2H to 2H' phase transition. In order to clearly distinguish the structures, filtered images painted with different color (Fig. S4e-S4h) are presented below each original ADF image. Fig. S4a shows the original 2H-MoS$_2$ lattice with two visible nonparallel α-phases (green ribbon in Fig. S4e). At the corner of two met α-phases, 28 Mo atoms in an equilateral triangle area (~6.2 nm$^2$) gliding in the distance which is equivalent to $a/\sqrt{3}$ (the lattice constant $a = 3.16$ Å) along the direction pointed by white arrows in Fig. S4e.

The Mo-plane gliding changes the orientation of 2H phase into opposite orientation (2H') as shown in the blue colored area in Fig. S4f. After Mo-plane gliding, the β-boundary which consists of four-coordinated S can be found between 2H and 2H' phases as a twin boundary (Fig. S4b and yellow ribbon in Fig. S4f). Note here that, the α-phase (3 zigzag chains) in the bottom of Fig. S4e donates 6 Mo atoms for gliding and the remains two zigzag chains transform to γ-boundary (bottom purple ribbon in Fig. S4f). On the other hand, the α-phase on the top of Fig. S4e accepts 6 Mo atoms and transform to two γ-boundaries sitting side-by-side. Fig. S4g illustrates the Mo atoms sequentially glide at β-boundary along the direction pointed by arrows, and then the 2H phase increases with β-boundary migration. The β-boundary migrates rightward consecutively and also a non-straight boundary part pointed by yellow arrow (Fig. S4h).

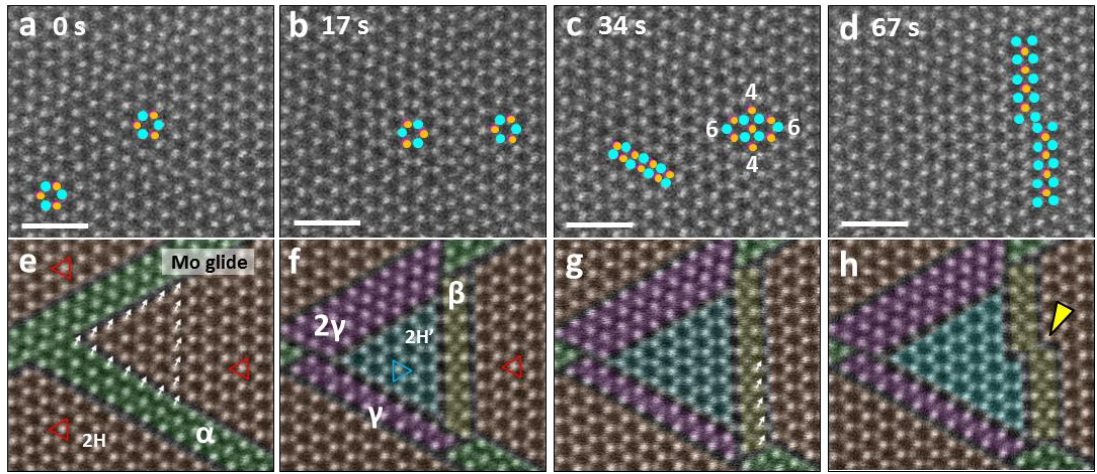

**Fig. S4. (a)-(d)** Sequential ADF images of MoS$_2$ phase transition from 2H to 2H' by Mo plane gliding. **(e)-(f)** The filtered images painted with color to distinguish boundary structures in advanced.

## 3.2. 2H→2H' and 2H→1T' phase transitions without intermediate state (α-phase).

We have observed over 100 sequential images and 98% of the studies have found that the intermediate state (α-phase) formed before phase transformations. The very rare cases that proceeded phase transition without the assistance of α-phase are also found in the 2H to 2H' and 2H to 1T phase transitions as shown in Figs. S5a-S5d (Movie M7) and Figs. S5e-S5h (Movie M8), respectively. Here, the 2H→2H' transition requires two S planes gliding simultaneously. On the other hand, the 2H→1T transition requires only one S plane gliding. In both cases, all of the boundaries between two phases are γ-boundaries pointed by green arrows in Fig. S5. Without the supporting from α-phase, these structures are not relatively unstable which expanding in size faster and also can disappear in a sudden.

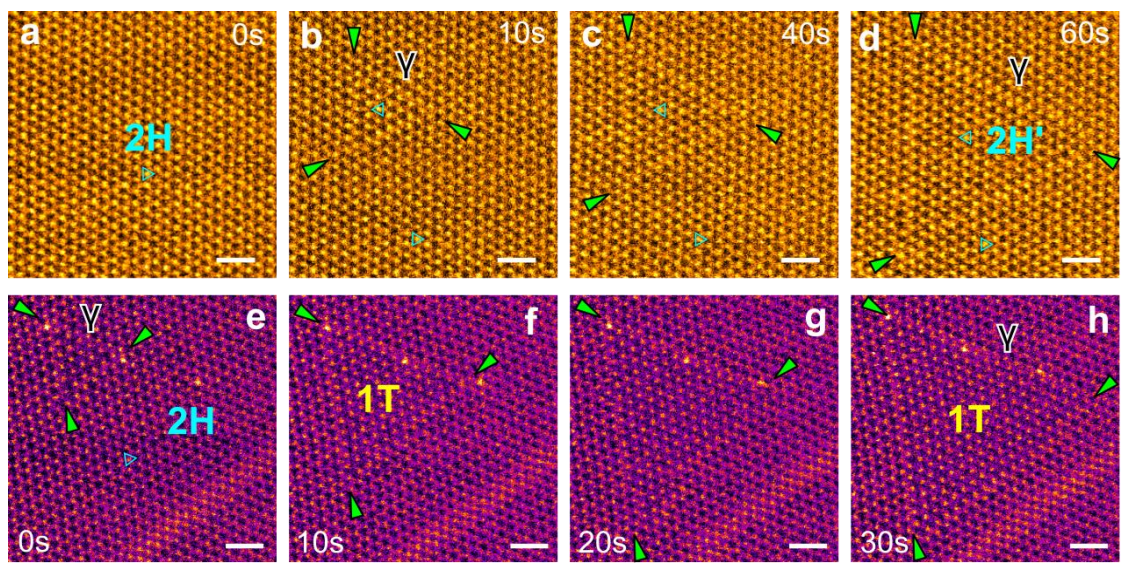

**Fig. S5**. **(a)-(d)** Sequential ADF images of 2H to 2H' phase transition without α-phase forming in advance. Two S planes glide simultaneously and from only γ-boundary at three edges. Boundaries are pointed by greed arrows. **(e)-(h)** 2H to 1T phase transition without α-phase by gliding only one S plane.

### 3.3. Atomic gliding models for 2H↔2H' and 2H↔1T phase transitions.

Figure S6 summarized the atomic gliding models for 2H↔2H' and 2H↔1T phase transitions. The transition is reversible between two phases. Fig. S6a represents the 2H to 2H' phase transition by S+S' gliding simultaneously and forms a γ-boundary between two phases. Fig. S6b represents the 2H to 2H' phase transition by Mo gliding and forms a β-boundary between two phases. Fig. S6c represents the 2H to 1T phase transition by Mo+S(or S') gliding simultaneously and forms a β-boundary between two phases. Fig. S6d represents the 1T to 2H phase transition by S(or S') gliding and forms a γ-boundary between two phases.

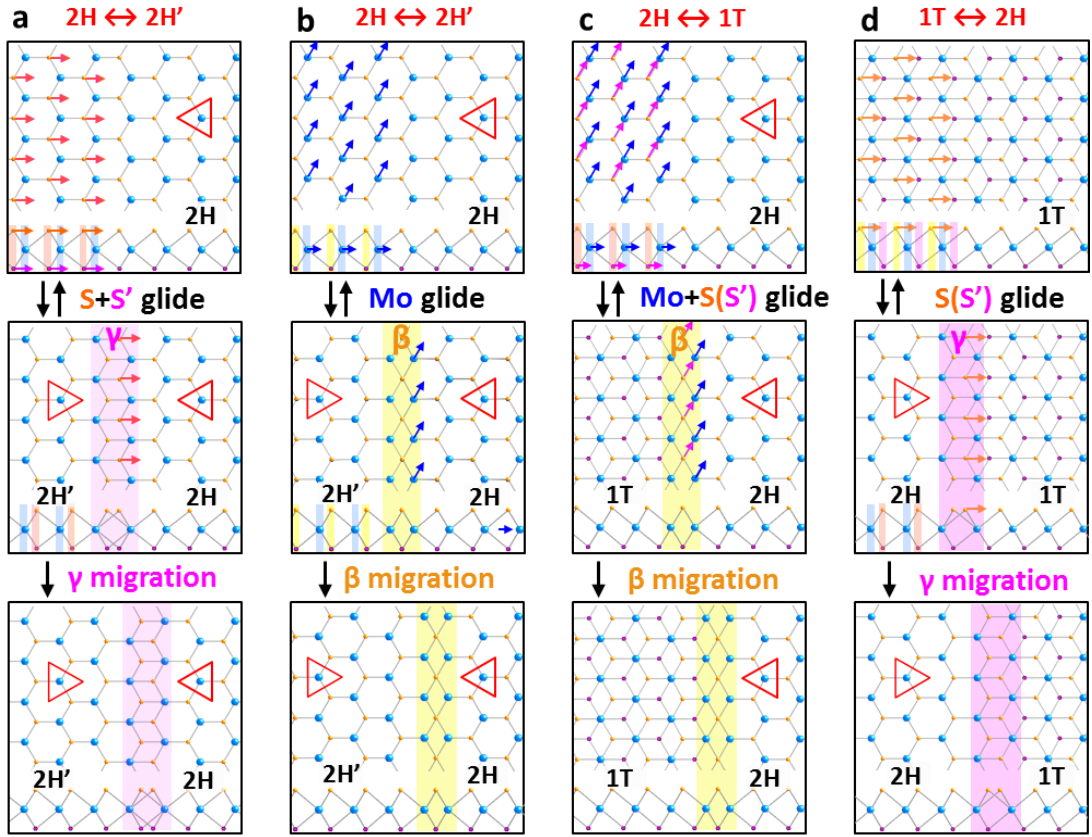

**Fig. S6.** Atomic gliding models of phase transitions. **(a)** 2H↔2H' transition by S+S' gliding. **(b)** 2H↔2H' transition by Mo gliding. **(c)** 2H↔1T transition by Mo+S(or S') gliding. **(d)** 1T↔2H transition by S(or S') gliding.

**Reference.**

[1] Komsa, H. P. *et al.* Two-dimensional transition metal dichalcogenides under electron irradiation: Defect production and doping. *Phys. Rev. Lett.* **109**, 035503 (2012).